\documentclass[12pt]{article}
\usepackage{jeosman}
\usepackage{epsfig}
\usepackage{amsmath}
\usepackage{amssymb}

\newcommand{\nnn}{{\boldsymbol n}}
\newcommand{\rrr}{{\boldsymbol r}}
\newcommand{\rrrp}{{\boldsymbol r}_\perp^{{}}}

\newcommand{\EEE}{{\boldsymbol E}}
\newcommand{\DDD}{{\boldsymbol D}}
\newcommand{\nablabf}{{\boldsymbol \nabla}}
\newcommand{\Laplace}{{\boldsymbol \nabla}_\perp^2}
\newcommand{\LaplaceThreeD}{{\boldsymbol \nabla}^2}

\begin{document}

\title{Photonic crystal fibres:\\ mapping Maxwell's equations onto a Schr\"odinger equation eigenvalue problem}

\author{Niels Asger Mortensen}

\address{MIC -- Department of Micro and Nanotechnology, NanoDTU,\\Technical
University of Denmark, DK-2800 Kongens Lyngby, Denmark}

\email{nam@mic.dtu.dk}


\begin{abstract}
We consider photonic crystal fibres (PCFs) made from arbitrary
base materials and introduce a short-wavelength approximation
which allows for a mapping of the Maxwell's equations onto a
dimensionless eigenvalue equations which has the form of the
Schr\"odinger equation in quantum mechanics. The mapping allows
for an entire analytical solution of the dispersion problem which
is in qualitative agreement with plane-wave simulations of the
Maxwell's equations for large-mode area PCFs. We offer a new angle
on the foundation of the endlessly single-mode property and show
that PCFs are endlessly single mode for a normalized air-hole
diameter smaller than $\sim 0.42$, independently of the base
material. Finally, we show how the group-velocity dispersion
relates simply to the geometry of the photonic crystal cladding.
\end{abstract}

\maketitle

\section{Introduction}

\begin{figure}[t!]
\begin{center}
\epsfig{file=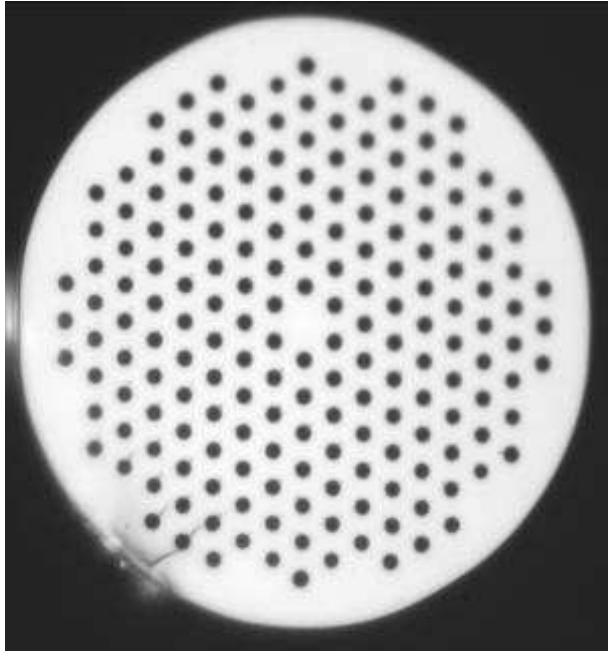,
width=0.5\columnwidth,clip,angle=0}
\end{center}
\caption{Optical micrograph of a silica-based PCF with a pitch
$\Lambda$ of order 10 microns and $d/\Lambda\sim 0.5$. Courtesy of
Crystal Fibre A/S, www.crystal-fibre.com. } \label{fig1}
\end{figure}

Photonic crystal fibres (PCF) are a special class of optical
fibres where a remarkable dielectric topology facilitates unique
optical properties. Guiding of light is accomplished by a highly
regular array of air holes running along the full length of the
fibre, see Fig.~\ref{fig1}. In their simplest form the fibres
employ a single dielectric base material (with dielectric function
$\varepsilon_b=n_b^2$) and for the majority of fabricated fibres
silica has been the most common
choice~\cite{knight1996,birks1997}. This preference has of course
been highly motivated by silica's transparency in the visible to
infrared regimes, but has also been strongly driven by the highly
matured glass technology used in the fabrication of standard
telecommunication fibres. However, a growing interest in fibre
optics for other wavelengths and light sources has recently led to
a renewed interest in other base materials including chalcogenide
glass~\cite{monro2000b}, lead silicate glass~\cite{kumar2002a},
telluride glass~\cite{kumar2003}, bismuth
glass~\cite{ebendorff-heidepriem2004}, silver
halide~\cite{rave2004}, teflon~\cite{goto2004a}, and
plastics/polymers~\cite{vaneijkelenborg2001}.

The huge theoretical and numerical effort in the PCF community has
been important in understanding the dispersion and modal
properties, but naturally there has been a clear emphasize on
silica-based PCFs. PCFs fabricated from different base materials
typically share the same overall geometry with the air holes
arranged in a triangular lattice and the core defect being formed
by the removal of a single air-hole. However, it is not yet fully
clarified how the established understanding may be transferred to
the development of PCFs made from other base materials. To put it
simple; PCFs made from different base materials share the same
topology, but to which extend do they also have optical
characteristics in common?

When addressing this question it is important to realize that the
otherwise very useful concept of scale-invariance of Maxwell's
equations~\cite{joannopoulos} is of little direct use in this
particular context. Since PCFs made from different base materials
do not relate to each other by a linear scaling of the dielectric
function, $\varepsilon(\rrr)\nrightarrow s^2\varepsilon(\rrr)$,
scale invariance cannot be applied directly to generalize the
results for e.g. silica to other base materials. Electro-magnetic
perturbation theory is of course one possible route allowing us to
take small changes of the base-material refractive index into
account, but in this paper we will discuss and elaborate on an
alternative approach which was put forward
recently~\cite{mortensen2005a}.

As a starting point we take the classical description based on
Maxwell's equations, which for linear dielectric materials
simplify to the following vectorial wave-equation for the
electrical field~\cite{joannopoulos}
\begin{equation}\label{eq:EEE}
\nablabf\times\nablabf\times\EEE(\rrr) =
\varepsilon(\rrr)\frac{\omega^2}{c^2}\:\EEE(\rrr).
\end{equation}
Here, $\varepsilon$ is the dielectric function which in our case
takes the material values of either air or the base material. For
a fibre geometry with $r_\parallel$ along the fibre axis we have
$\varepsilon(\rrr)=\varepsilon(\rrrp)$ and as usual we search for
solutions of the plane-wave form $e^{i(\beta r_\parallel-\omega
t)}$ with the goal of calculating the dispersion $\omega(\beta)$
relation.

\section{Short-wavelength approximations}

Today, several approaches are available for solving
Eq.~(\ref{eq:EEE}) numerically, including plane-wave, multi-pole,
and finite-element
methods~\cite{johnson2001,kuhlmey2002b,saitoh2002}. Such methods
will typically be preferred in cases where highly quantitative
correct results are called for. However, numerical studies do not
always shine much light on the physics behind so here we will take
advantage of several approximations which allow for analytical
results in the short-wavelength limit $\lambda\ll\Lambda$. The key
observation allowing for the approximations is that typically the
base material has a dielectric function exceeding that of air
significantly, $\varepsilon_b\gg 1$. In that case it is well-known
that the short-wavelength regime is characterized by having the
majority of the electrical field residing in the high-index base
material while the fraction of electrical field in the air holes
is vanishing.

The, at first sight very crude, approximation is simply to neglect
the small fraction of electrical power residing in the air-hole
regions. Mathematically the approximation is implemented by
imposing the boundary condition that $\EEE$ is zero at the
interfaces $\partial\Omega$ to air. Since the displacement field
$\DDD=\varepsilon\EEE$ is divergence free we have $0=\varepsilon
\nablabf\cdot\EEE+ \EEE\cdot\nablabf \varepsilon \approx
\varepsilon \nablabf\cdot\EEE$ and the wave
equation~(\ref{eq:EEE}) now reduces to
\begin{equation}\label{eq:EEEapprox1}
-\LaplaceThreeD\EEE(\rrr) \approx
\varepsilon_b\frac{\omega^2}{c^2}\:\EEE(\rrr),\quad
\EEE(\rrr)\Big|_{\rrr\in\partial\Omega}=0,
\end{equation}
and since solutions are of the plane-wave form $e^{i(\beta
r_\parallel-\omega t)}$ along the fiber axis we get
\begin{equation}\label{eq:EEEapprox}
-\Laplace\EEE(\rrrp)+ \beta^2\EEE(\rrrp) \approx
\varepsilon_b\frac{\omega^2}{c^2}\:\EEE(\rrrp),\quad
\EEE(\rrrp)\Big|_{\rrrp\in\partial\Omega}=0,
\end{equation}
where $\Laplace$ is two-dimensional Laplacian in the transverse
plane. The essentially scalar nature of this equation makes the
problem look somewhat similar to the more traditional scalar
approximation
\begin{equation}\label{eq:scalar1}
\Laplace\Psi(\rrrp) + \varepsilon(\rrrp)\frac{\omega^2}{c^2}
\:\Psi(\rrrp)\approx \beta^2\:\Psi(\rrrp),
\end{equation}
that has been applied successfully to PCFs in the short-wavelength
regime~\cite{riishede2003a}. However, while that approach took the
field and the dielectric function in the air holes into account we
shall here solve Eq.~(\ref{eq:EEEapprox}) with the boundary
condition that $\EEE$ is zero at the interfaces to air. We note
that this of course is fully equivalent to solving
Eq.~(\ref{eq:scalar1}) with the boundary condition that $\Psi$ is
zero at the interfaces to air. Obviously, the scalar problem posed
by Eq.~(\ref{eq:EEEapprox}) is separable and formally we have that
\begin{equation}\label{eq:omega(beta)1}
\omega=\sqrt{\Omega_{\perp}^2+\Omega_\parallel^2}=\sqrt{\Omega_{\perp}^2+\left(c\beta/n_b\right)^2}
\end{equation}
where $\Omega_\parallel=c\beta/n_b$ is the frequency associated
with the longitudinal plane-wave propagation with a linear
dispersion relation (i.e. $\Omega_\parallel\propto \beta$) and
$\Omega_{\perp}$ is the frequency associated with the transverse
confinement/localization. At this point we already note how an
arbitrary base-material refractive index $n_b$ enters in the
frequency associated with the longitudinal plane-wave propagation.
This is solely possible because we have a Dirichlet boundary
condition in Eq.~(\ref{eq:EEEapprox}) which does not depend on the
refractive index of the base material.

\section{The Schr\"odinger equation eigenvalue problem}\label{sec:schrodinger}

In the following we rewrite Eq.~(\ref{eq:omega(beta)1}) as
\begin{equation}\label{eq:omega(beta)}
\omega=\frac{c}{n_b}\sqrt{\gamma^2\Lambda^{-2}+\beta^2}
\end{equation}
where now $\gamma$ is a dimensionless number characterizing the
confinement/localization. It is of purely geometrical origin and
thus only depends on the normalized air-hole diameter $d/\Lambda$.
From Eq.~(\ref{eq:EEEapprox}) it follows that $\gamma$ is an
eigenvalue governed by a scalar two-dimensional Schr\"{o}dinger
equation
\begin{equation}\label{eq:schroding}
-\Lambda^2\Laplace\psi(\rrrp)=\gamma^2\psi(\rrrp),\quad
\psi(\rrrp)\Big|_{\rrrp\in\partial\Omega}=0.
\end{equation}
The same equation was recently studied in work on anti-dot
lattices in a two-dimensional electron gas~\cite{Flindt:2005} and
there will be many similarities between such quantum mechanical
electron systems and the present electromagnetic problem. In the
language of quantum mechanics the wavefunction $\psi$ is subject
to hard-wall boundary conditions corresponding to an infinite
potential barrier in the air-hole regions. The task of calculating
the optical dispersion properties has now reduced to solving the
scalar two-dimensional Schr\"odinger-like eigenvalue problem posed
by Eq.~(\ref{eq:schroding}). The strength is obviously that
Eq.~(\ref{eq:schroding}) is material independent and thus we may
in principle solve it once and for all to get the geometry
dependence of $\gamma$ and thereby the optical dispersion of the
PCF. In the following we solve Eq.~(\ref{eq:schroding}) for
different classes of modes and compare approximate results to
corresponding numerically exact finite-element solutions.

\subsection{The fundamental space-filling mode}

The fundamental space-filling mode is a key concept in
understanding the light guiding properties of PCFs. It is the
fundamental de-localized mode of Eq.~(\ref{eq:schroding}) in an
infinite periodic structure and the corresponding mode index
corresponds to an effective material index of the artificial
periodic photonic crystal lattice. In the language of band
diagrams and Brillouin zones it is defined in the $\Gamma$-point
where Bloch's theorem is particular simple; as illustrated in
Fig.~\ref{fig3} the wave function is subject to a simple Neumann
boundary condition on the edge $\partial\tilde\Omega$ of the
Wigner--Seitz cell, i.e.
\begin{equation}\label{eq:schroding_FSM}
-\Lambda^2\Laplace\psi(\rrrp)=\gamma_{\rm cl} ^2\psi(\rrrp),\quad
\psi(\rrrp)\Big|_{\rrrp \in\partial\Omega}= \nnn \cdot \nablabf
\psi(\rrrp)\Big|_{\rrrp\in\partial\tilde\Omega}=0,
\end{equation}
where $\nnn$ is a unit vector normal to the boundary
$\partial\tilde\Omega$. We thus have to solve an eigenvalue
problem on a doubly-connected domain. By a conformal mapping,
leaving the Laplacian and the boundary conditions invariant, one
may in principle transform the hexagonal shape with the circular
hole of diameter $d$ into a simple annular region of inner
diameter $d$ and outer diameter $2R$~\cite{laura:1972} leaving us
with an eigenvalue problem in the radial coordinate. Finding the
exact conformal mapping may be a complicated task so here we
simply use the approximate mapping with
\begin{equation}
R= \Lambda \sqrt{\frac{\sqrt{3}}{2\pi}}\simeq 0.53\, \Lambda
\end{equation}
which serves to conserve the area of the Wigner--Seitz unit cell.
Neglecting any small changes to the eigenvalue equation caused
during the mapping we get the following eigenvalue equation in the
radial coordinate,
\begin{equation}\label{eq:schroding_cl}
-\Lambda^2(\partial_r^2+r^{-1}\partial_r)\psi(r)\approx
\gamma_{\rm cl} ^2\psi(r),\quad \psi(d/2)=\psi'(R)=0.
\end{equation}
The solution is given by a linear combination of the Bessel
functions $J_0$ and $Y_0$ and the eigenvalue is determined by the
roots of the following expression
\begin{equation}\label{eq:roots_cladding}
J_1\left(\frac{3^{\frac{1}{4}} }{{\sqrt{2\,\pi }}}\gamma_{\rm
cl}\right) Y_0\left(\frac{1}{2}\frac{d}{\Lambda} \gamma_{\rm
cl}\right)=
  J_0\left(\frac{1}{2}\frac{d}{\Lambda} \gamma_{\rm cl}\right)Y_1\left(\frac{3^{\frac{1}{4}}\, }{{\sqrt{2\,\pi
}}}\gamma_{\rm cl}\right).
\end{equation}
In general the equation cannot be solved exactly except for at the
point
\begin{equation}\label{eq:claddingeq}
\left\{\frac{d}{\Lambda},\gamma_{\rm
cl}^2\right\}=\left\{2\frac{3^{1/4}}{\sqrt{2\pi
}}\:\frac{\alpha_{0,1}}{\alpha_{1,1}},
\frac{2\pi}{3^{1/2}}\:\alpha_{1,1}^2\right\}\simeq
\left\{0.66,53.26\right\},
\end{equation}
as is easily verified by insertion and using that $\alpha_{n,m}$
is the $m$th zero of the $n$th Bessel function, i.e.
$J_n(\alpha_{n,m})=0$. In particular we have that
$\alpha_{0,1}\simeq 2.405$ and $\alpha_{1,1}\simeq 3.8317$.
However, expanding the left and right-hand sides of
Eq.~(\ref{eq:claddingeq}) around this point to first order in both
$\gamma_{\rm cl}$ and $d/\Lambda$ we get an equation from which we
may isolate $\gamma_{\rm cl}$ yielding an approximate analytical
solution of the form
\begin{equation}\label{eq:gammacl}
\gamma_{\rm cl}^2\simeq \left({\cal C}_1  + \frac{{\cal
C}_2}{{\cal C}_3 - d/\Lambda} \right)^2.
\end{equation}
The coefficients are given by expressions involving Bessel
functions $Y_0$ and $Y_1$ evaluated at $\alpha_{0,1}$ and
$\alpha_{1,1}$, see Appendix, but for simplicity we here only give
the corresponding numerical values of these expressions; ${\cal
C}_1\simeq -0.2326$, ${\cal C}_2\simeq 2.7040$, and ${\cal
C}_3\simeq 1.0181 $.

\subsection{Core-defect modes}

\begin{figure}[t!]
\begin{center}
\epsfig{file=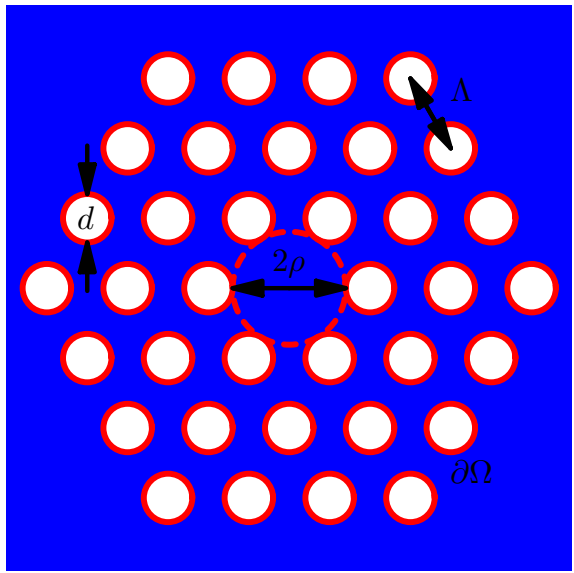,
width=0.5\columnwidth,clip,angle=0}
\end{center}
\caption{The topology of a PCF formed by omitting an air hole of
diameter $d$ in a triangular lattice of air holes with pitch
$\Lambda$. Dirichlet boundary conditions are applied to the
boundary $\partial\Omega$ indicated by red solid lines. The red
dashed line indicates twice an effective core radius
$\rho=\Lambda-d/2$.} \label{fig2}
\end{figure}

\begin{figure}[t!]
\begin{center}
\epsfig{file=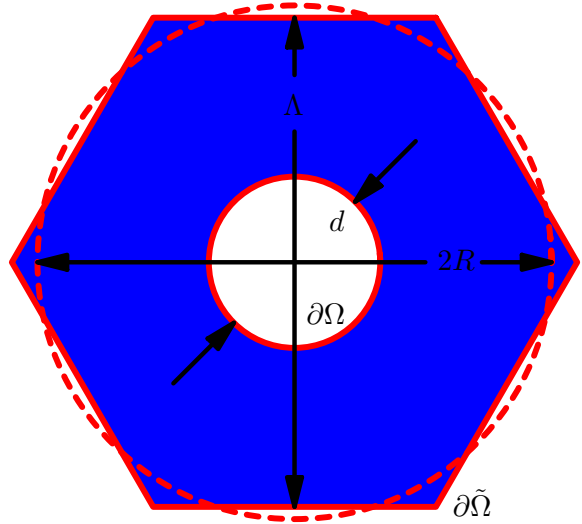,
width=0.5\columnwidth,clip,angle=0}
\end{center}
\caption{The hexagonal Wigner--Seitz unit cell of the periodic
triangular lattice with lattice constant/pitch $\Lambda$. The
dashed line indicates the annular region which has the same area
as the doubly-connected domain of hexagonal outer shape. Dirichlet
and Neumann boundary conditions apply to the inner and outer
boundaries $\partial\Omega$ and $\partial\tilde\Omega$,
respectively. } \label{fig3}
\end{figure}

While the fundamental space-filling mode, discussed above, is a
de-localized mode it is also possible to have strongly localized
states, especially in the presence of defects such as one formed
by removing an air hole from the otherwise periodic lattice of air
holes, see figures~\ref{fig1} and \ref{fig2}. In this way light
can be guided along the defect thus forming the core of the fibre.
The requirement for corralling the light is of course that the
core-defect mode has an eigenvalue $\gamma_{c,1}^2$ not exceeding
the corresponding eigenvalue $\gamma_{\rm cl}^2$ of the
surrounding photonic crystal cladding. To put it slightly
different the defect in the air-hole lattice effectively
corresponds to a potential well of depth $\gamma_{\rm cl}^2$ and a
radius roughly given by $\rho=\Lambda-d/2$, see Fig.~\ref{fig2}.

\begin{figure}[t!]
\begin{center}
\epsfig{file=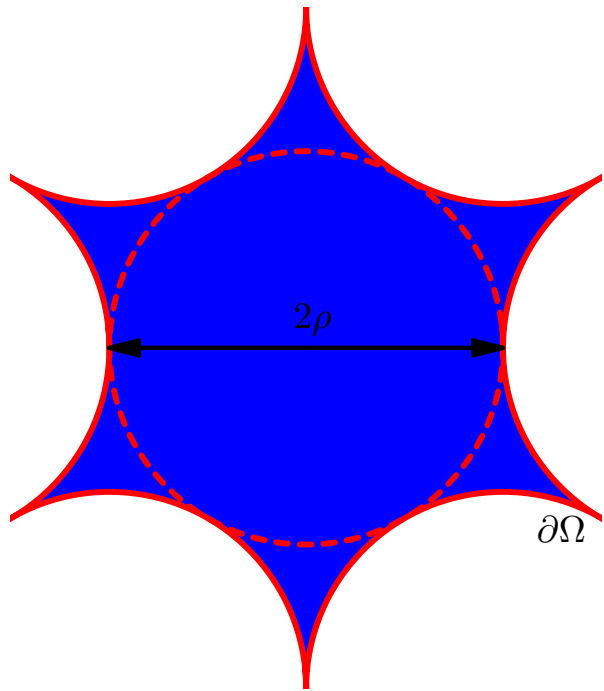,
width=0.5\columnwidth,clip,angle=0}
\end{center}
\caption{Zoom in on the core-defect in the extreme limit of
$d/\Lambda=1$. A Dirichlet boundary condition applies to the
boundary $\partial\Omega$ and the dashed circle indicates the
approximate core radius $\rho=\Lambda-d/2$. } \label{fig4}
\end{figure}

In order to calculate the fundamental core-defect mode we first
note that in the limit of $d/\Lambda$ approaching 1, the problem
can be approximated with that of a two-dimensional spherical
infinite potential well with radius $\rho=\Lambda-d/2$, see
Fig.~\ref{fig4}. For this problem the lowest eigenvalue is
\begin{equation}
\gamma_{c,1}^2\sim
\left(\frac{\Lambda}{\rho}\right)^2\alpha_{0,1}^2=\frac{\alpha_{0,1}^2}{\left(1-\frac{1}{2}\frac{d}{\Lambda}\right)^2}.
\end{equation}
Although this expression yields the correct scaling with
$d/\Lambda$, the approximation obviously breaks down for small
values of $d/\Lambda$. In that limit we follow the ideas of
Glazman \emph{et al.}~\cite{Glazman:1988}, who studied quantum
conductance through narrow constrictions. The effective
one-dimensional energy barrier for transmission through the
constriction of width $\Lambda-d$ between two neighbouring holes
has a maximum value of $\pi^2$ in the limit where $d$ goes to
zero. We thus approximate the $d/\Lambda\rightarrow 0$ problem
with that of a two-dimensional spherical potential well of height
$\pi^2$ and radius $\Lambda$,
\begin{equation}\label{eq:schroding_c_low}
-\Lambda^2(\partial_r^2+r^{-1}\partial_r)\psi(r)+\pi^2
\Theta(r-\Lambda)\psi(r)= \tilde\gamma_{c,1} ^2\psi(r),
\end{equation}
where $\Theta(x)$ is the Heaviside step function which is zero for
$x<0$ and unity for $x>0$. The lowest eigenvalue
$\tilde\gamma_{c,1}^2$ for this problem is the first root of the
following equation
\begin{equation}\label{eq:pi2}
\tilde\gamma_{c,1}\frac{
      J_1(\tilde\gamma_{c,1})}
      {J_0(\tilde\gamma_{c,1})}
    = \sqrt{\pi^2 -
          \tilde\gamma_{c,1}^2}
\frac{K_1\left(\sqrt{\pi^2 - \tilde\gamma_{c,1}^2}\:\right)}
      {K_0\left(
      \sqrt{\pi^2 - \tilde\gamma_{c,1}^2}\:\right)}.
    \end{equation}
In the case where the potential $\pi^2$ is replaced by an infinite
potential the first root is given by $\alpha_{0,1}^2$. Lowering
the confinement potential obviously shifts down the eigenvalue and
in the present case the eigenvalue is roughly
$\tilde\gamma_{c,1}^2 \sim \pi$. In fact, expanding the equation
to first order in $\tilde\gamma_{c,1}$ around e.g. $\sqrt{\pi}$ we
get a straight forward, but long, expression with the numerical
value $\tilde\gamma_{c,1}^2\simeq 3.221$ which is in excellent
agreement with a numerical solution of the equation. Correcting
for the low-$d/\Lambda$ behaviour we thus finally get
\begin{equation}\label{eq:gammac1}
\gamma_{c,1}^2 \simeq
\tilde\gamma_{c,1}^2+\left\{1-\lim_{d/\Lambda\rightarrow
0}\right\}\frac{\alpha_{0,1}^2}{\left(1-\frac{1}{2}\frac{d}{\Lambda}\right)^2}=
\tilde\gamma_{c,1}^2+\frac{\left(4-\frac{d}{\Lambda}\right)\frac{d}{\Lambda}}{\left(2-\frac{d}{\Lambda}\right)^2}\:\alpha_{0,1}^2.
\end{equation}
For the first high-order mode we may apply a very similar
analysis. This mode has a finite angular momentum of $\pm 1$ with
a radial $J_1$ solution yielding
\begin{equation}\label{eq:gammac2}
\gamma_{c,2}^2 \simeq
\tilde\gamma_{c,2}^2+\frac{\left(4-\frac{d}{\Lambda}\right)\frac{d}{\Lambda}}{\left(2-\frac{d}{\Lambda}\right)^2}\:\alpha_{1,1}^2.
\end{equation}
Here, $\tilde\gamma_{c,2}^2\simeq 7.673$ is the second eigenvalue
of the two-dimensional spherical potential well of height $\pi^2$
and radius $\Lambda$. Again, it can be found from an equation very
similar to Eq.~(\ref{eq:pi2}).

\subsection{Two-dimensional finite-element solutions}

\begin{figure}[t!]
\begin{center}
\epsfig{file=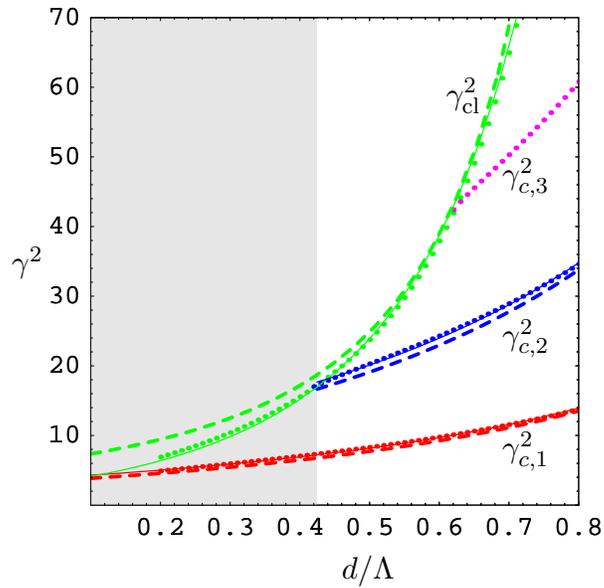,
width=0.5\columnwidth,clip,angle=0}
\end{center}
\caption{Geometrical eigenvalues $\gamma^2$ versus normalized
air-hole diameter $d/\Lambda$. The data points are results of
finite-element simulations while the corresponding dashed lines
are the approximate results in Eqs.~(\ref{eq:gammacl}),
(\ref{eq:gammac1}), and (\ref{eq:gammac2}). The thin solid lines
shows the numerical fits listed in Eqs.~(\ref{eq:fitc1}),
(\ref{eq:fitc2}), and (\ref{eq:fitcl}). Note how only a single
core-defect mode is supported in the shaded region,
$d/\Lambda\lesssim 0.42$. } \label{fig5}
\end{figure}

Above we have solved the geometrical eigenvalue problem
analytically by means of various approximations. In this section
we compare the quality of these results by direct comparison to a
numerically exact solution of the two-dimensional eigenvalue
problem. The developments in computational physics and engineering
have turned numerical solutions of partial differential equations
in the direction of a standard task. Here, we employ a
finite-element approach\cite{Comsol} to numerically solve
Eq.~(\ref{eq:schroding}) and calculate $\gamma^2$ versus
$d/\Lambda$. We employ an adaptive mesh algorithm to provide
efficient convergence. For cladding mode we implement
Eq.~(\ref{eq:schroding_FSM}) directly while for the defect modes
we solve Eq.~(\ref{eq:schroding}) on a finite domain of
approximate size $10\Lambda\times 10\Lambda$ with the defect
located in the center of the domain. For $d/\Lambda$ down to
around 0.1 we have found this to sufficient to adequate
convergence for the strongly localized defect states, thus
avoiding strong proximity from the domain-edge boundary condition
which for simplicity has been of the Dirichlet type.
Figure~\ref{fig5} summarizes our numerical results for the first
cladding mode $\gamma_{\rm cl}^2$ as well for the two first
core-defect modes $\gamma_{c,1}^2$ and $\gamma_{c,2}^2$. The
dashed lines indicate the corresponding approximate expressions in
Eqs.~(\ref{eq:gammacl}), (\ref{eq:gammac1}), and
(\ref{eq:gammac2}) obtained by analytical means. As seen the
qualitative agreement is excellent, but in order to facilitate
also quantitative applications of the results we also include the
results of numerical least-square error fits below. The thin solid
lines show the expressions
\begin{subequations}
\begin{align}
\gamma_{c,1}^2 &\simeq
3.666+\frac{\left(4-\frac{d}{\Lambda}\right)\frac{d}{\Lambda}}{\left(2-\frac{d}{\Lambda}\right)^2}\:\alpha_{0,1}^2,\label{eq:fitc1}\\
\gamma_{c,2}^2 &\simeq
8.691+\frac{\left(4-\frac{d}{\Lambda}\right)\frac{d}{\Lambda}}{\left(2-\frac{d}{\Lambda}\right)^2}\:\alpha_{1,1}^2,\label{eq:fitc2}\\
\gamma_{cl}^2 &\simeq \left(-2.82476  + \frac{5.23695}{1.17908 -
d/\Lambda} \right)^2\label{eq:fitcl},
\end{align}
\end{subequations}
which match the numerical data within a relative error of less
than $2\%$ around the most important cut-off region $d/\Lambda\sim
0.42$.

\section{Derived fiber optical parameters}

With Eq.~(\ref{eq:omega(beta)}) at hand we have now provided a
unified theory of the dispersion relation in the short-wavelength
regime for PCFs with arbitrary base materials and
Eq.~(\ref{eq:omega(beta)}) illustrates how geometrical confinement
modifies the linear free-space dispersion relation.

In fibre optics it is common to express the dispersion properties
in terms of the effective index $n_{\rm eff}=c\beta/\omega$ versus
the free-space wavelength $\lambda=c2\pi/\omega$. From
Eq.~(\ref{eq:omega(beta)}) it follows straightforwardly that
\begin{equation}\label{eq:neff}
n_{\rm
eff}=n_b\sqrt{1-\frac{\gamma^2}{4\pi^2n_b^2}\left(\frac{\lambda}{\Lambda}\right)^2}
\end{equation}
which obviously is in qualitative agreement with the accepted view
that $n_{\rm eff}$ increases monotonously with decreasing
wavelength and approaches $n_b$ in the asymptotic short-wavelength
limit as reported for {\it e.g.} silica-based
PCFs~\cite{birks1997}. Similarly, the group-velocity
$v_g=\partial\omega/\partial\beta$ becomes
\begin{equation}
v_g =
\frac{c}{n_b}\sqrt{1-\frac{\gamma_{c,1}^2}{4\pi^2n_b^2}\left(\frac{\lambda}{\Lambda}\right)^2}
\end{equation}
while the phase velocity $v_p=\omega/\beta$ becomes
\begin{equation}
v_p =
\frac{c}{n_b}\frac{1}{\sqrt{1-\frac{\gamma_{c,1}^2}{4\pi^2n_b^2}\left(\frac{\lambda}{\Lambda}\right)^2}}.
\end{equation}
The group-velocity dispersion is most often quantified by the
wave-guide dispersion parameter $D_{\rm
wg}=\partial(1/v_g)/\partial\lambda$ which becomes
\begin{equation}
D_{\rm wg} = \frac{1}{c\Lambda} \frac{\gamma_{c,1}^2}{4\pi^2 n_b}
\left[1-\frac{\gamma_{c,1}^2}{4\pi^2n_b^2}\left(\frac{\lambda}{\Lambda}\right)^2\right]^{-3/2},
\end{equation}
where any possible material dispersion $n_b(\lambda)$ has been
neglected. We note that since $D_{\rm wg}>0$ the geometry of the
air-hole lattice always causes a positive wave-guide dispersion
parameter. Large-mode area PCFs belong to the regime with
$\lambda\ll\Lambda$ were we predict the following general
magnitude of the wave-guide dispersion of a large-mode area PCF
\begin{equation}
\lim_{\lambda\ll\Lambda}D_{\rm wg} = \frac{1}{c\Lambda}
\frac{\gamma_{c,1}^2}{4\pi^2 n_b}.
\end{equation}
Finally, the recently suggested parameter $V_{\rm PCF}=\Lambda
\sqrt{\beta_{\rm cl}^2-\beta_{c,1}^2}$~\cite{mortensen2003c} can
be shown to be a purely geometrically defined parameter in the
large-mode area limit. From Eq.~(\ref{eq:omega(beta)}) it follows
straightforwardly that
\begin{equation}
\lim_{\lambda\ll\Lambda}V_{\rm PCF}=\sqrt{\gamma_{\rm
cl}^2-\gamma_{c,1}^2}.
\end{equation}
This implies that the endlessly single-mode
property~\cite{birks1997,mortensen2002a,mortensen2003c,Saitoh:2005}
is a wave phenomena independent of the base material refractive
index of purely geometrical origin. Higher-order modes are only
supported for $d/\Lambda\gtrsim 0.42$, as seen in Fig.~\ref{fig5},
for which $V_{\rm PCF}\gtrsim\pi$~\cite{mortensen2003c}.

\section{Comparison to fully-vectorial plane-wave simulations}

\begin{figure}[t!]
\begin{center}
\epsfig{file=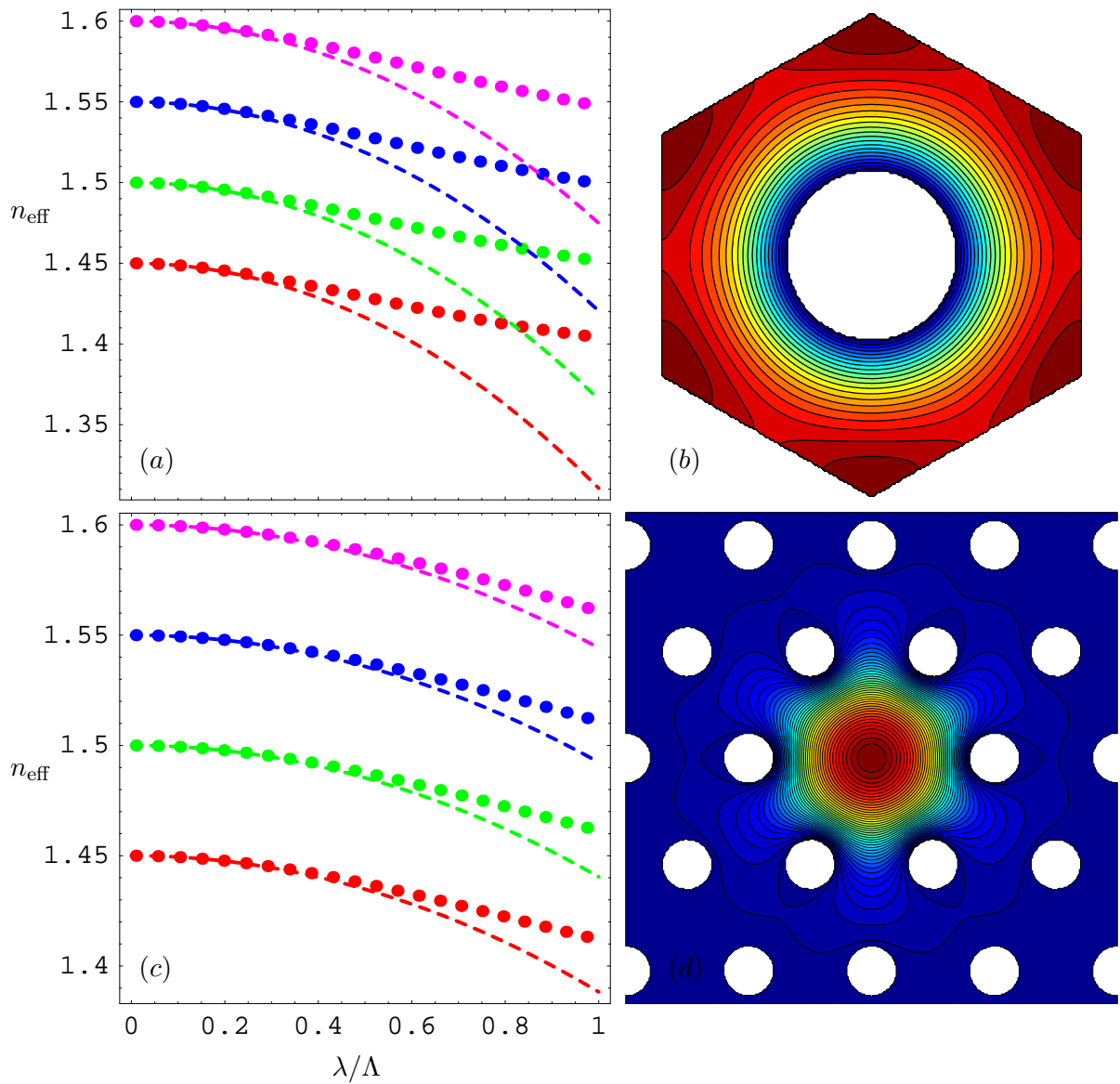, width=\columnwidth,clip,angle=0}
\end{center}
\caption{Effective index versus normalized wavelength
$\lambda/\Lambda$ for the fundamental space-filling mode (a) and
the fundamental core-defect mode (c) for a PCF with
$d/\Lambda=0.4$ with a base material refractive index $n_b$
varying from 1.45 to 1.6 in steps of 0.05 from below. The data
points are the results of a vectorial plane-wave simulation of
Eq.~(\ref{eq:EEE}) while the dashed lines show the corresponding
results based on Eq.~(\ref{eq:neff}) with Eqs.~(\ref{eq:fitc1})
and (\ref{eq:fitcl}) for geometrical eigenvalues $\gamma_{\rm
cl}^2$ and $\gamma_{c,1}^2$. Panels (b) and (d) show the
corresponding eigenfunctions $\psi$.} \label{fig6}
\end{figure}

In the previous sections substantial analytical progress was made,
but we still remain to address the question to which extend the
basic assumption behind Eq.~(\ref{eq:EEEapprox}) holds except that
it becomes exact for the fundamental modes as $\lambda/\Lambda$
approaches zero. In Fig.~\ref{fig6} we compare the analytical
results for the effective index in Eq.~(\ref{eq:neff}) to results
obtained by fully-vectorial plane-wave
simulations~\cite{johnson2001} of Eq.~(\ref{eq:EEE}). For the
fundamental space-filling mode we have employed a basis of
$2^6\times 2^6$ plane waves while for the core-defect modes we
have employed a super-cell configuration of size $10\Lambda\times
10\Lambda$ in the case of $d/\Lambda=0.4$.

Panel (a) shows results for the fundamental space-filling mode for
various values of the base refractive index. As clearly seen the
theory agrees excellently in the short-wavelength limit while
pronounced deviations occur as the wavelength increases and the
field penetrates deeper into the air hole regions and also
vectorial effects become important as reported
in~\cite{riishede2003a}. Similar conclusion applies to the
corresponding results in panel (c) for the fundamental core-defect
mode. While $d/\Lambda=0.4$ is a case of particular technological
interest because of its endlessly single-mode property we would
like to emphasize that equivalent agreement is found for other
values of $d/\Lambda$ (results not shown). However, since the
success of the approximation in Eq.~(\ref{eq:EEEapprox}) really is
that $\lambda/n_b\ll 2\rho=2\Lambda-d$ and $\lambda/n_b\ll
\Lambda-d$ for the core and cladding results, respectively, there
will of course be a small $d$-dependence. However, this also
indicates that the agreement will increase with an increasing
refractive index $n_b$ of the base material as has also been
observed for even higher values of $n_b$ than those studied in
Fig.~\ref{fig6} (results not shown).

\section{Discussion and conclusion}

It is today more than 10 years ago that the PCF was invented by
Russell and co-workers~\cite{knight1996} and there is a still
growing community of researchers directing their efforts toward
fabrication and experimental studies of silica-based PCFs as well
as quantitative modelling of the optical properties. At the same
time the community is obviously facing the challenges and
opportunities of new exciting fiber materials. PCFs made from
different base materials share the same topology so it seems quite
natural to assume that they also have at least some basic optical
properties in common. Let's return to the question posed in the
introduction: to which extend do PCFs made from different
materials have optical characteristics in common? The present work
do to some extend address this question. Most importantly we
illustrate how the waveguide dispersion originates from the
geometrical transverse confinement/localization of the mode and
how the endlessly single-mode property arises as a sole
consequence of the geometry which acts as a modal sieve. In
particular we have shown how PCFs are endlessly single-mode for
$d/\Lambda\lesssim 0.42$ irrespectively of the base material an
analytical expression for the wave-guide dispersion parameter
applicable to large-mode area fibres.

For small-core PCFs our theory provides qualitative correct
results though more quantitative insight still calls for fully
vectorial simulations, see e.g.~\cite{saitoh:2006a}. However, for
large-mode area PCFs our expressions not only gives qualitative
insight, but also quantitative correct expressions which may be
used in straightforward design of large-mode area PCFs with
special properties with respect to the group-velocity dispersion
and the susceptibility of the fundamental mode with respect to
longitudinal non-uniformities~\cite{mortensen2003b}.

\section*{Acknowledgments}

C. Flindt, J. Pedersen, and A.-P. Jauho are acknowledged for
stimulating discussions on the results in
Section~\ref{sec:schrodinger} and for sharing numerical results.

\appendix

\section{Taylor expansion of Eq.~(\ref{eq:roots_cladding})}

Taylor expanding the left and right-hand sides of
Eq.~(\ref{eq:roots_cladding}) around the point
\begin{equation}
\left\{d_0\:,\:\gamma_0\right\}\equiv\left\{2\frac{3^{1/4}}{\sqrt{2\pi
}}\:\frac{\alpha_{0,1}}{\alpha_{1,1}}\:\Lambda \:,\:
\sqrt{\frac{2\pi}{3^{1/2}}}\:\alpha_{1,1}\right\}
\end{equation}
to first order in both $\gamma_{\rm cl}$ and $d/\Lambda$ we get
\begin{align}
0=& \left(\frac{d-d_0}{d_0}\right)+  \frac{
       Y_1^2(\alpha_{1,1})-Y_0^2(\alpha_{0,1})  }{
      Y_1^2(\alpha_{1,1})}\left(\frac{\gamma_{\rm cl} -\gamma_0
  }{\gamma_0}\right)\nonumber\\
  &+
     \frac{
     Y_1(\alpha_{0,1})}
      { Y_0(\alpha_{0,1})} \frac{d_0}{\Lambda}\left(\frac{d-d_0}{d_0}\right)\left(\gamma_{\rm cl} -\gamma_0
      \right)\nonumber\\
      &+{\cal O}\left([d-d_0]^2\right)+{\cal O}\left([\gamma_{\rm
      cl}-\gamma_0]^2\right).
\end{align}
Next, solving for $\gamma_{\rm cl}$ we may write the result in the
form of Eq.~(\ref{eq:gammacl}) with
\begin{subequations}
\begin{align}
{\cal C}_1 &=  \gamma_0 - \frac{ Y_0(\alpha_{0,1})}{
Y_1(\alpha_{0,1})}\left(\frac{d_0}{\Lambda}\right)^{-1},\\
{\cal C}_2&= \frac{Y_0^4(\alpha_{0,1}) - Y_0^2(\alpha_{0,1})\,
Y_1^2(\alpha_{1,1})}
  {\gamma_0\,{Y_1^2(\alpha_{0,1})}\,{
  Y_1^2(\alpha_{1,1})}}\left(\frac{d_0}{\Lambda}\right)^{-1},\\
{\cal C}_3&= \frac{{Y_0^3(\alpha_{0,1})} -
     Y_0(\alpha_{0,1})\,{Y_1^2(\alpha_{1,1})}}{\gamma_0\,
     Y_1(\alpha_{0,1})\,{Y_1^2(\alpha_{1,1})}}+
     \frac{d_0}{\Lambda},
\end{align}
\end{subequations}
which have the numerical values listed in the paragraph below
Eq.~(\ref{eq:gammacl}).

\newpage


\begin{thebibliography}{10}
\newcommand{\enquote}[1]{``#1''}

\bibitem{knight1996}
J.~C. Knight, T.~A. Birks, P.~S.~J. Russell, and D.~M. Atkin,
  \enquote{All-silica single-mode optical fiber with photonic crystal
  cladding}, Opt. Lett. \textbf{21} 1547--1549 (1996).

\bibitem{birks1997}
T.~A. Birks, J.~C. Knight, and P.~S.~J. Russell,
\enquote{Endlessly single mode
  photonic crystal fibre}, Opt. Lett. \textbf{22} 961--963 (1997).

\bibitem{monro2000b}
T.~M. Monro, Y.~D. West, D.~W. Hewak, N.~G.~R. Broderick, and
D.~J. Richardson,
  \enquote{Chalcogenide holey fibres}, Electron. Lett. \textbf{36} 1998--2000
  (2000).

\bibitem{kumar2002a}
V.~V. R.~K. Kumar, A.~K. George, W.~H. Reeves, J.~C. Knight,
P.~S.~J. Russell,
  F.~G. Omenetto, and A.~J. Taylor, \enquote{Extruded soft glass photonic
  crystal fiber for ultrabroad supercontinuum generation}, Opt. Express
  \textbf{10} 1520--1525 (2002).

\bibitem{kumar2003}
V.~V. R.~K. Kumar, A.~K. George, J.~C. Knight, and P.~S.~J.
Russell,
  \enquote{Tellurite photonic crystal fiber}, Opt. Express \textbf{11}
  2641--2645 (2003).

\bibitem{ebendorff-heidepriem2004}
H.~Ebendorff-Heidepriem, P.~Petropoulos, S.~Asimakis, V.~Finazzi,
R.~C. Moore,
  K.~Frampton, F.~Koizumi, D.~J. Richardson, and T.~M. Monro, \enquote{Bismuth
  glass holey fibers with high nonlinearity}, Opt. Express \textbf{12}
  5082--5087 (2004).

\bibitem{rave2004}
E.~Rave, P.~Ephrat, M.~Goldberg, E.~Kedmi, and A.~Katzir,
\enquote{Silver
  halide photonic crystal fibers for the middle infrared}, Appl. Opt.
  \textbf{43} 2236--2241 (2004).

\bibitem{goto2004a}
M.~Goto, A.~Quema, H.~Takahashi, S.~Ono, and N.~Sarukura,
\enquote{Teflon
  photonic crystal fiber as terahertz waveguide}, Jap. J. Appl. Phys.
  \textbf{43} L317--L319 (2004).

\bibitem{vaneijkelenborg2001}
M.~A. {van Eijkelenborg}, M.~C.~J. Large, A.~Argyros, J.~Zagari,
S.~Manos,
  N.~A. Issa, I.~Bassett, S.~Fleming, R.~C. McPhedran, C.~M. {de Sterke}, and
  N.~A.~P. Nicorovici, \enquote{Microstructured polymer optical fibre}, Opt.
  Express \textbf{9} 319--327 (2001).

\bibitem{joannopoulos}
J.~D. Joannopoulos, R.~D. Meade, and J.~N. Winn, \emph{Photonic
crystals:
  molding the flow of light} (Princeton University Press, Princeton, 1995).

\bibitem{mortensen2005a}
N.~A. Mortensen, \enquote{Semianalytical approach to
short-wavelength
  dispersion and modal properties of photonic crystal fibers}, Opt. Lett.
  \textbf{30} 1455 -- 1457 (2005).

\bibitem{johnson2001}
S.~G. Johnson and J.~D. Joannopoulos, \enquote{Block-iterative
frequency-domain
  methods for \uppercase{M}axwell's equations in a planewave basis}, Opt.
  Express \textbf{8} 173--190 (2001).

\bibitem{kuhlmey2002b}
B.~T. Kuhlmey, T.~P. White, G.~Renversez, D.~Maynstre, L.~C.
Botton, C.~M. {de
  Sterke}, and R.~C. McPhedran, \enquote{Multipole method for microstructured
  optical fibers. II. Implementation and results}, J. Opt. Soc. Am. B
  \textbf{19} 2331--2340 (2002).

\bibitem{saitoh2002}
K.~Saitoh and M.~Koshiba, \enquote{Full-vectorial
imaginary-distance beam
  propagation method based on finite element scheme: Application to photonic
  crystal fibers}, IEEE J. Quantum Electron. \textbf{38} 927--933 (2002).

\bibitem{riishede2003a}
J.~Riishede, N.~A. Mortensen, and J.~L{\ae}gsgaard, \enquote{A
poor man's
  approach to modelling of microstructured optical fibers}, J. Opt. A: Pure.
  Appl. Opt. \textbf{5} 534 (2003).

\bibitem{Flindt:2005}
C.~Flindt, N.~A. Mortensen, and A.~P. Jauho, \enquote{Quantum
computing via
  defect states in two-dimensional antidot lattices}, Nano Lett. \textbf{5}
  2515 -- 2518 (2005).

\bibitem{laura:1972}
P.~A. Laura, E.~Romanelli, and M.~J. Maurizi, \enquote{On the
analysis of
  waveguides of double-connected cross-section by the method of conformal
  mapping}, J. Sound Vibr. \textbf{20} 27--38 (1972).

\bibitem{Glazman:1988}
L.~I. Glazman, G.~K. Lesovik, D.~E. Khmelnitskii, and R.~I.
Shekter,
  \enquote{Reflectionsless quantum transport and fundamental
  ballistic-resistance steps in microscopic constrictions}, JETP Lett.
  \textbf{48} 238--241 (1988).

\bibitem{Comsol}
Femlab, http://www.comsol.com.

\bibitem{mortensen2003c}
N.~A. Mortensen, J.~R. Folkenberg, M.~D. Nielsen, and K.~P.
Hansen,
  \enquote{Modal cut-off and the $V$--parameter in photonic crystal fibers},
  Opt. Lett. \textbf{28} 1879--1881 (2003).

\bibitem{mortensen2002a}
N.~A. Mortensen, \enquote{Effective area of photonic crystal
fibers}, Opt.
  Express \textbf{10} 341--348 (2002).

\bibitem{Saitoh:2005}
K.~Saitoh, Y.~Tsuchida, M.~Koshiba, and N.~A. Mortensen,
\enquote{Endlessly
  single-mode holey fibers: the influence of core design}, Opt. Express
  \textbf{13} 10833 -- 10839 (2005).

\bibitem{saitoh:2006a}
K.~Saitoh, M.~Koshiba, and N.~A. Mortensen, \enquote{Nonlinear
photonic crystal
  fibres: pushing the zero-dispersion toward the visible}, Special issue on nanophotonics to appear in New J. Phys.
  (2006). http://arxiv.org/physics/0608142

\bibitem{mortensen2003b}
N.~A. Mortensen and J.~R. Folkenberg, \enquote{Low-loss criterion
and effective
  area considerations for photonic crystal fibers}, J. Opt. A: Pure Appl. Opt.
  \textbf{5} 163--167 (2003).

\end{thebibliography}

\newpage

\end{document}